\documentstyle{article}

\def\tr{{\rm Tr}}
\def\C{{\cal C}}
\def\Hol{{\rm Hol}}
\def\tr{{\rm Tr}}
\def\Link{{\rm Link}}
\def\sLink{{\rm sLink}}

\def\be{\begin{equation}}
\def\ee{\end{equation}}
\def\bea{\begin{eqnarray}}
\def\eea{\end{eqnarray}}

\def\bb#1{\hbox{\mybb#1}}

\def\zet{\bb{Z}}

\def\R4{\bb{R}^4}
\font\mybb=msbm10 at 12pt

\def\tr{{\rm Tr}}
\def\C{{\cal C}}
\def\Hol{{\rm Hol}}
\def\tr{{\rm Tr}}
\def\Link{{\rm Link}}
\def\sLink{{\rm sLink}}

\def\unita{{1 \kern-.30em 1}}

\begin{document}

\begin{center}
 
{\large \bf  A NEW NONPERTURBATIVE APPROACH TO QCD  BY BF THEORY}\\ 
  
\vspace{0.5cm}
\centerline{\bf Maurizio Martellini$^\dagger$, Mauro Zeni} 
{\sl Dipartimento di Fisica, Universit\`a di Milano \\
and \\  
I.N.F.N. \ - \  Sezione di Milano, \\
Via Celoria 16 \ \ 20133 \ Milano \ \ ITALY}
\centerline{\sl $^\dagger$Landau Network  at ``Centro Volta'', Como, ITALY}
\vskip .2cm
{\bf Francesco Fucito}
\vskip 0.1cm
\centerline{\sl Dipartimento di Fisica, Universit\`a di Roma II ``Tor Vergata"}
\centerline{\sl and}  
\centerline{\sl I.N.F.N. \ - \  Sezione di Roma II, }
\centerline{\sl Via Della Ricerca Scientifica \ \ 00133 \ Roma \ \ ITALY}
\end{center}
\abstract{
Yang-Mills theory in the first order formalism appears as the deformation
of a topological field theory, the pure BF theory. In this approach
new non local observables are inherited from the topological theory
and the operators entering the t'Hooft algebra find an explicit 
realization. A calculation of the {\it vev}'s of these operators
is performed in the Abelian Projection gauge.
}
\vfill

\section{First order formalism}
The first order form of pure euclidean Yang-Mills theory 
is described by the action functional \cite{fmz,ccgm}
\be 
S_{BF-YM} = \int_{M^4} d^4x {i\over 2}\varepsilon^{\mu\nu\alpha\beta}
B^a_{\mu\nu}F^a_{\alpha\beta}
+g^2 \int_{M^4} d^4x B^a_{\mu\nu}B^{a\mu\nu} \quad ,
\label{uno.1}
\ee  
with the usual definition
$F^a_{\mu\nu}\equiv 2\partial_{[\mu}A^a_{\nu ]} +f^{abc}A^b_\mu A^c_\nu $ 
and where $B$ is a Lie algebra valued 2-form; 
the usual gauge symmetry takes the form 
$\delta A_\mu = D_\mu c$, $\delta B_{\mu\nu}= -[c,B_{\mu\nu}]$ 
with $D_\mu\equiv\partial_\mu -i[A_\mu ,\cdot ]$.
The first term in the r.h.s. of (\ref{uno.1}) is the action of a 
topological quantum field theory, the so-called BF theory \cite{blau}; indeed 
the action (\ref{uno.1}) acquires a ``topological'' symmetry iff $g=0$.
In this case
\be 
\delta A^a_\mu = 0 \qquad ,\qquad 
\delta B^a_{\mu\nu} = 
 2\partial_{[\mu}\psi^a_{\nu ]}  +2f^{abc}A^b_{[\mu} \psi^c_{\nu ]}\quad ,
\label{uno.3}
\ee 
where  $\psi $ is a 1-form ghost. Note that due to zero modes 
$\delta \psi_\mu =D_\mu\phi $ and $\delta\phi =0$ in 
the transformations (\ref{uno.3}). This symmetry is then reducible, 
allowing for a ghosts of ghosts 
structure. 
The second term in r.h.s. of (\ref{uno.1}) is an explicit symmetry breaking 
for the topological symmetry (\ref{uno.3}) and restores a local gauge dynamics.
The field equations of (\ref{uno.1}) read  
\be
*F^a_{\mu\nu}={ig^2\over 2}B^a_{\mu\nu}\qquad ,\qquad 
\varepsilon ^{\mu\nu\alpha\beta}D_\nu B^a_{\alpha\beta} =0\quad ,
\label{uno.4}
\ee
where 
$*F_{\mu\nu}\equiv\frac 12 \varepsilon_{\mu\nu\alpha\beta} 
F^{\alpha\beta}$ 
is the dual field strength. Note that on-shell  $B$ coincides with 
the dual field strength and satisfies the Bianchi identities. This is no longer 
true off-shell. In this case the quantum theory may 
introduce monopoles charges through $B$ which  
is not constrained by a Bianchi identity.
These topologically non-trivial magnetic configurations should enter the YM 
vacuum and be in correspondence with a deformation of the instanton sector. 
By 
performing a formal perturbation theory in $g\neq 0$ around the 
topological BF theory (the topological embedded sector in YM) it is easily 
seen that 
the BF topological vacuum in an axial gauge, say $B^-=0$, is given by the 
quasi anti-self-dual instanton configurations
$F^+_{\mu\nu} =o(g)$. 

\section{A new observable}
A new non-local observable associated to an orientable surface 
$\Sigma\in M^4$  is naturally introduced in the BF formulation 
of QCD \cite{fmz,ccgm}, 
\be                       
M(\Sigma ,C)\equiv \tr \exp \{ik
\int_{\Sigma } d^2y 
\ \Hol^y_{\bar x} (\gamma ) B(y) \Hol_y^{\bar x} (\gamma^{\prime} )\} \quad ,
 \label{due.1} 
\ee 
where $\Hol_{\bar x}^y(\gamma )$ denotes the  holonomy 
along the open path $\gamma\equiv\gamma_{\bar x y}$ with initial and final 
point $\bar x$ and $y$ respectively,
\be 
\Hol_{\bar x}^y(\gamma )\equiv P\exp \{i\int_{\bar x}^y dx^{\mu} A_{\mu}(x)\}
\quad .
\label{due.2}
\ee 
In (\ref{due.1}) $k$ is an arbitrary expansion parameter, 
$\bar x$ is a {\it fixed} point (we do not integrate over $\bar x$) 
and the relation between the assigned paths 
$\gamma$, $\gamma^{\prime}$ over $\Sigma$ and the closed contour $C$ is the 
following: $C$ 
starts from the fixed point $\bar x$, connects a point $y\in\Sigma$ by 
the open path $\gamma_{\bar x y}$ and then returns back to the
neighborhood of $\bar x$ by 
$\gamma_{y\bar x }^{\prime}$, 
(which is not restricted to coincide with the inverse 
$(\gamma_{\bar x y})^{-1}=\gamma_{y\bar x }$). From the neighborhood 
of $\bar x$ the path starts again to connect another point 
$y^{\prime}\in\Sigma$. Then it 
returns back to the neighborhood of $\bar x$ and so on until all point
on $\Sigma$ are connected. The path $C$ is generic and 
no particular ordering prescription is required. Of course the 
quantity (\ref{due.1}) is path dependent and our strategy is to regard it as 
a loop variable once the surface $\Sigma$ is given. 

Our main result is that the observable $M$ gives an explicit analytic 
realization of the `t Hooft loop operator \cite{thooft1} 
(the dual variable to the 
Wilson loop).
This may be realized using $3+1$-hamiltonian formalism. 
When considering the action of $M(C)$ on a physical state 
$|A>$ we find \cite{fmz}, in the fundamental representation, 
\be 
 M(C)|A(\vec x)>\simeq \tr \{ e^{ig\Phi_C}\unita\} |A(\vec x)> \quad ,
\label{due.3} 
\ee 
where $\Phi_C\equiv {2\pi k(N^2-1)\over gN}\sLink (C)$. The self-linking 
$\sLink (C)$ is an integer defined as the linking between the curve $C$ and 
its framing contour $C^\prime$, the latter being a point splitting 
regularization of the former, i.e. if $C\equiv \{ x^i(t)\}$ then 
$ C^{\prime}\equiv \{ y^i(t)=x^i(t)+ 
\epsilon n^i(t)\}$ with $ \epsilon >0$ and 
$n^i$ a versor orthogonal to $C$. 
$\sLink (C) \equiv \Link(C,C^{\prime})_{\epsilon\rightarrow 0}$ 
equals the number of windings of $C^{\prime}$ around $C$. 
Thus with a proper choice of the parameter $k=4\pi /(N^2 -1)$ 
the operator $M(C)$ generates 
multivalued (i.e. singular) gauge 
transformations in the center of the group $SU(N)$; this is the 
defining property of the `t Hooft magnetic operator.
The vacuum expectation values of the operator $M(C)$ and of the Wilson loop 
$W(C)$ label the phases of the theory; in the confinement phase, $M(C)$ 
and $W(C)$
are expected to develop a perimeter and an area law respectively.
\section{Computation of $<M>$}
We compute the {\it vev} of $M$ in the abelian 
projection  gauge \cite{thooft2}, which in our case is better implemented 
choosing
$B$ in the adjoint representation of $SU(N)$.
In this approach a monopole dynamics is supposed to arise and dominate 
\cite{thooft2,eza} the YM vacuum, discarding the ``charged'' \cite{thooft2} 
gauge degrees of freedom. Indeed in this approach we find \cite{fmz}  
magnetic abelian configurations $\bar\alpha$ of the gauge field satisfying 
a monopole equation 
\be 
\frac 12 \varepsilon_{\mu\nu}^{\ \ \alpha\beta}\partial_\alpha 
\bar\alpha^i_\beta ={4\pi q\over N}\omega_{\Sigma\mu\nu}
\cos (g\oint_C\bar\alpha_i)\quad ,
\label{tre.3}  
\ee 
$\omega_{\Sigma}$ is a closed form and may be chosen 
self-dual, i.e. $*\omega_{\Sigma}=\omega_{\Sigma}$. Locally it can be written 
as $\omega_{\Sigma}\simeq\delta^2(\Sigma )$. 
The saddle point evaluation for $<M>$  \cite{fmz} gives (in form language 
$\omega_{\Sigma}=\omega_{\Sigma\mu\nu}dx^\mu\otimes   dx^\nu$)
\be 
{<M(C)>\over <1>}\simeq N\exp \{-{2\pi^2 n^2\over g^2}\sum_i
\int 
\omega_{\Sigma}\wedge\omega_{\Sigma^{\prime}}
\cos (g\oint_C\bar\alpha_i )
\cos (g\oint_{C^{\prime}}\bar\alpha_i^{\prime} )\}\quad ,
\label{tre.2} 
\ee 
where $n\in \zet$. The flux of the magnetic solutions of (\ref{tre.3}) 
satifies a proper Dirac quantization 
$ g\oint_C\bar\alpha_i ={4\pi gq\over N} = 2\pi n$, with 
$n\in\zet$.In this case, the quantity 
$\int \omega_{\Sigma}\wedge\omega_{\Sigma^{\prime}}$ in (\ref{tre.2}), 
where $\Sigma^{\prime}$($\supset C^{\prime}$) 
is a point splitting regularization of $\Sigma$($\supset C$), 
equals $\sLink (C)$ \cite{horouno,horo}. 
In a lattice regularization 
it is found that $\sLink (C)\sim {L(C)\over a}$, where $L$ is the 
perimeter of $C$ measured in latting spacing $a$.  To summarize, in 
the continuum limit we find
the expected perimeter law for the `t Hooft operator
\be 
{<M(C)>\over <1>}\sim \exp \{-{8\pi^2 n^2N\over 4g^2}{L(C)\over l}\}\quad ,
\label{tre.4} 
\ee 
$l$ is the magnetic vortex penetration lenght. 
It can be shown \cite{fmz} that 
one-loop quantum effects amount to the replacement of the coupling $g$ in 
(\ref{tre.4}) with the renormalized one, i.e. $8\pi /g^2_R =8\pi /g^2_\mu
-\beta_1\ln ({\sqrt {(-p^2)}\over \mu})$, where $p$ is the momentum scale, 
$\mu$ is the subtraction point and $\beta_1=\frac {11}3 N$. We also have 
$1/l\to \Lambda_{QCD}$ with $\Lambda_{QCD}$ some typical physical mass scale 
in QCD.
\section{Computation of $<W>$ and conclusions}
We consider now the computation of the {\it vev} of the Wilson loop, which 
in the confinement phase of QCD is expected to develop an area law. We start 
by rewriting it in terms of the non abelian Stokes theorem \cite{ara} 
\be
W(\C) \equiv W(\Sigma ,C)=Tr P_{\Sigma}
\exp \{ i\int_{\Sigma} \Hol_{\bar x}^x (\gamma )
F(x) \Hol^{\bar x}_x(\gamma^{\prime})\}  \quad ,
\label{qua.1} 
\ee
where $\C=\partial\Sigma$ and $P_{\Sigma}$ means surface path ordering.
Using the functional identity 
$  
F^a_{\rho\sigma}(x) \exp \{ -\frac i 4 \int *B^a_{\mu\nu}F^a_{\mu\nu} \}$ 
$=4i {\delta } / {\delta *B^a_{\rho\sigma}(x)} \exp \{ -\frac i 4 \int 
*B^a_{\mu\nu}F^a_{\mu\nu}\}
$ 
we obtain \cite{fmz} the ``duality relation''
\be 
<W(\C)>=<M^*(\Sigma ,C)>\equiv 
<Tr P_{\Sigma} \exp \{ -{g^2\over 2}\int_{\Sigma}
\Hol_{\bar x}^x (\gamma )
*B(x) \Hol^{\bar x}_x(\gamma^{\prime})\} > \quad .
\label{qua.2}
\ee 
To calculate (\ref{qua.2}) we expand $M^*$ in $g$; the first relevant 
contraction encountered at lower level is given in terms
of $<A*B>$, involving the off
diagonal propagator $<AB>$ present both in the BF-YM theory \cite{mz} 
as well as in the pure BF theory. Therefore we find 
\be 
{<M^*(\C)>\over <1>}  =e^{-{g^2\over 2} c_2(t)\oint_{C}\int_{\Sigma}<A*B>}
\Delta (\Sigma)\quad ,
\label{qua.3}   
\ee 
where $c_2(t)$ is the 2th Casimir of the group representation  
and $\Delta (\Sigma)$ depends 
on higher order integrations over $\Sigma$. 
It may be shown \cite{fmz} that 
\be 
\oint_{C}\int_{\Sigma}<A*B>\sim \int_{\Sigma} Link(C,\Sigma^*_x)\quad , 
\label{qua.4}
\ee 
where $Link(C,\Sigma^*_x)$ is the linking number between the curve $C$ and the 
dual plane $\Sigma^*_x$ in $x$ to $\Sigma$ \cite{horouno,horo}.
Using again a lattice regularization, (13) is found to be
proportional to the area of $\Sigma$. We then obtain
\be 
{<W(\C)>\over <1>}\sim \exp \{-\sigma(l) A(\Sigma)\}\quad , 
\label{qua.5}
\ee 
where $\sigma(l)$ is the string tension defined at tree level by 
\be 
\sigma(l) \equiv g^2 ({N^2-1\over 16N})
\frac 1{l^2}\quad .
\label{qua.6} 
\ee
Again quantum effect should amount to the 
replacement of the bare coupling with the running one 
and also $\sigma$ with the renormalization group invariant quantity 
$\bar\sigma (\Lambda_{QCD})$. A rough estimate of $\bar\sigma$ 
for energy scales between $10$ and 
$100(Gev)^2$ and $\Lambda_{QCD}\sim 0.5$~$Gev$ gives
$
\bar\sigma \sim 1.2\times 0.25 (Gev)^2 
$, 
which is of the same order of magnitude of the experimental value. 

In conclusion, 
in the framework of the first order (BF) formalism for YM theory we have 
introduced an explicit analytic representation for the `t Hooft algebra, using 
non local operators whose {\it vev} are naturally given in terms of 
geometrical quantities like linking numbers between curves and surfaces. The 
underlying dynamics is understood in terms of monopole vortex lines which 
should enter the theory by means of the auxiliary field $B$, thus supporting 
the picture of the dual superconductor vacuum. The explicit calculation 
produces perimeter and area laws for the operators $M(C)$ and $W(C)$,  
corresponding to the expected confining phase for the theory.

This work has been partially supported by Ministero dell'Universit\`a e della 
Ricerca Scientifica e Tecnologica and by E.E.C. Grants CHRX-CT92-0035 and 
SC1$^*$-CI92-0789.


\section*{References}
\begin{thebibliography}{99}

\bibitem{fmz} F.~Fucito, M.~Martellini and M.~Zeni, 
``{\it The BF formalism for QCD and Quark Confinement}'', hep-th/9605018. 

\bibitem{ccgm}
A.~S.~Cattaneo, P.~Cotta-Ramusino, A.~Gamba and M.~Martellini, 
{\it Phys. Lett.} {\bf B355} (1995) 245; A.~S.~Cattaneo, P.~Cotta-Ramusino,
J.~Fr\"ohlich and M.~Martellini, {\it J. Math. Phys.} {\bf 36} (1995) 6137.

\bibitem{blau}
M.~Blau and G.~Thompson, {\it Ann. Phys.}  {\bf 205} (1991) 130; 
N.~Maggiore and S.~P.~Sorella, {\it Int. J. Mod. Phys.} {\bf A8} (1993) 929.

\bibitem{thooft1}
G.~`t~Hooft, {\it Nucl.Phys.}  {\bf B138} (1978) 1; {\bf B153} (1979) 141.
			     
\bibitem{thooft2}
G.~`t~Hooft, {\it Nucl. Phys.}  {\bf B190}[FS3] (1981) 455. 

\bibitem{eza}
Z.~F.~Ezawa~and~A.~Iwazaki,~{\it Phys.Rev.}~{\bf D25}~(1982)~2681;
~{\it Phys.Rev.}{\bf D26}(1982)~631.

\bibitem{horouno}
G.~Horowitz, {\it Comm.Math.Phys.}{\bf 125}(1989) 417.

\bibitem{horo}
G.~T.~Horowitz and M.~Srednicki, {\it Comm. Math. Phys.}  {\bf 130} 
(1990) 83.

\bibitem{ara}
I.~Ya~Araf'eva, {\it Theor. Math. Phys.}  {\bf 43} (1980) 353.

\bibitem{mz}
M.~Martellini and M.~Zeni, 
``{\it Diagrammatic Feynman rules and $\beta$-function 
for the BF approach to QCD}'', and F.~Fucito, M.~Martellini,
A.~Tanzini and M.~Zeni, ``{\it The Topological Embedding of the BF
Theory in Yang-Mills}", in preparation. 

\vfill\eject 


\end{thebibliography}
\end{document}